\documentclass[runningheads]{llncs}
\usepackage{graphicx}

\usepackage{amsmath}
\usepackage{amssymb}

\usepackage{booktabs}
\usepackage{multirow}
\usepackage{multicol}
\usepackage{makecell}
\usepackage{bbding}

\usepackage{xcolor}
\begin{document}
\title{End-to-end Multi-source Visual Prompt Tuning for Survival Analysis in Whole Slide Images}

\author{Zhongwei Qiu\inst{1,4,5}\and
Hanqing Chao\inst{1,4} \and 
Wenbin Liu\inst{2} \and
Yixuan Shen\inst{2} \and
Le Lu\inst{1} \and
Ke Yan\inst{1,4} \and
Dakai Jin\inst{1} \and
Yun Bian\inst{2} \and
Hui Jiang\inst{3}
}
%index{Zhang, Yejia}
%index{Chao, Hanqing}
%index{Qiu, Zhongwei}
%index{Liu, Wenbin}
%index{Shen, Yixuan}
%index{Sapkota, Nishchal}
%index{Gu, Pengfei}
%index{Chen, Danny Z.}
%index{Bian, Yun}
%index{Jiang, Hui}
%index{Lu, Le}
%index{Yan, Ke}
%index{Jin, Dakai}

\institute{
    DAMO Academy, Alibaba Group \\
    \email{\{hanqing.chq, qiuzhongwei.qzw\}@alibaba-inc.com}  
    \and
    Departments of Radiology, \\
  Changhai Hospital, Shanghai, China \\
    \and
    Departments of Pathology, \\
  Changhai Hospital, Shanghai, China \\
    \and
    Hupan Lab, Hangzhou, China
    \and
    Zhejiang University, Hangzhou, China
}

\authorrunning{Z. Qiu et al.}
\titlerunning{End-to-End Multi-Source VPTSurv in WSIs}
% \author{Paper 1677}
% % \authorrunning{F. Author et al.}
% \institute{Anonymous Institutions}
\maketitle              % typeset the header of the contribution
\begin{abstract}
Survival analysis using pathology images poses a considerable challenge, as it requires the localization of relevant information from the multitude of tiles within whole slide images (WSIs). Current methods typically resort to a two-stage approach, where a pre-trained network extracts features from tiles, which are then used by survival models. This process, however, does not optimize the survival models in an end-to-end manner, and the pre-extracted features may not be ideally suited for survival prediction.
To address this limitation, we present a novel end-to-end Visual Prompt Tuning framework for survival analysis, named VPTSurv. VPTSurv refines feature embeddings through an efficient encoder-decoder framework. The encoder remains fixed while the framework introduces tunable visual prompts and adaptors, thus permitting end-to-end training specifically for survival prediction by optimizing only the lightweight adaptors and the decoder.
Moreover, the versatile VPTSurv framework accommodates multi-source information as prompts, thereby enriching the survival model. VPTSurv achieves substantial increases of 8.7\% and 12.5\% in the C-index on two immunohistochemical pathology image datasets. These significant improvements highlight the transformative potential of the end-to-end VPT framework over traditional two-stage methods.

\keywords{Survival Analysis \and Prompt Tuning \and End-to-end \and WSI.}
\end{abstract}
\section{Introduction}
Survival analysis aims to predict the risk of death events occurring in patients at a given time point. Accurate survival prediction can provide critical information for physicians in diagnostic and treatment planning. Modeling the tumor microenvironment within pathology images has been proven to be important for achieving prognostic analysis~\cite{abduljabbar2020geospatial,kuroda2021tumor}. Many works have focused on extracting useful information from WSIs~\cite{shao2021transmil,chen2021whole,lu2021data,chen2022scaling}, as well as integrating histology knowledge with biological pathways for the purpose of survival prediction~\cite{chen2021multimodal,zhou2023cross,Xu_2023_ICCV}. However, despite the remarkable progress, efficiently pinpointing critical data for the challenging task of survival prediction from gigapixel-scale WSIs remains a significant hurdle.

To facilitate survival analysis on WSIs, the mainstream paradigm adopts a two-stage pipeline: 1) offline feature extraction of millions of patches sliced from the WSI to reduce computational demands, followed by 2) feature aggregation with algorithms like multiple instance learning (MIL)~\cite{chikontwe2020multiple,shao2021transmil,lu2021data,chen2021whole,shao2023hvtsurv} to learn task-specific features.
Such a paradigm can be encapsulated within an encoder-decoder framework, with a frozen encoder for feature extraction. The main focuses of current work are on developing strategies for patch aggregations to boost the decoder's capability. 
For example, the MIL scheme formulates each WSI as a bag and forms multiple aggregation models~\cite{sharma2021cluster,wang2019weakly} or embedding learning models~\cite{ilse2018attention,lu2021data} to learn bag-level representations. Recently, attention-based MIL strategies such as DeepAttnMISL~\cite{yao2020whole}, TransMIL~\cite{shao2021transmil}, and HVTSurv~\cite{shao2023hvtsurv} have shown significant progress due to their ability to model long-term context with attention mechanisms~\cite{li2021dual,ilse2018attention}.

The aforementioned methods liberate the process of handling WSIs, thereby allowing a focus on the design of the decoder component. However, these methods exhibit an inherent performance bottleneck: their performance ceiling is limited by the features extracted offline.
In this paper, we argue that the features extracted by common-used pre-trained networks~\cite{lu2021data,huang2023visual} are not well-optimized for survival analysis, while end-to-end training can learn more suitable feature representation for survival prediction.
Nevertheless, end-to-end training of a stronger extractor on WSIs requires GPUs with huge memory sizes, which is not practical for most researchers. The reasons are the enormous numbers of patches in WSIs and the gradient computation of the extractor.
Moreover, even if end-to-end training is achieved through large-scale distributed training or model partitioning techniques, the scarcity of annotated prognostic data (such as a few hundred cases) may also lead to model non-convergence or severe overfitting.

To tackle the above issues, we introduce visual prompt tuning (VPT) into survival analysis and propose a novel end-to-end multi-source VPT framework for survival analysis, named VPTSurv. VPTSurv formulates the survival prediction as a unified encoder-decoder pipeline and performs end-to-end training. 
To overcome the challenges of end-to-end training, VPTSurv freezes the parameters of the pre-trained image encoder while adding learnable visual prompts and adaptors to tune the feature embedding for the survival decoder. This mechanism reduces the learnable parameters and keeps the capacity of the original pre-trained encoder. 
VPTSurv is an extensible framework that accommodates a small number of patches during training while permitting a larger number of patches during inference. These mechanisms enable lightweight and efficient end-to-end training for survival prediction.
Furthermore, VPTSurv is a unified framework that leverages the VPT to incorporate multi-source information as prompts, enhancing the performance of survival prediction.

Our contributions are threefold: 1) we present the first end-to-end VPT framework for survival analysis in WSIs, 2) we introduce a novel multi-source VPT mechanism that utilizes diverse information as prompts to improve survival prediction, and 3) we offer new insights into the scheme for processing WSIs.

% \begin{itemize}
%     \item We propose the first end-to-end trainable framework for survival analysis in WSI, which brings a new insight into the process of WSI.
%     \item We propose the first visual prompt tuning model for survival analysis in WSI.
%     \item We propose a cross-scale 
% \end{itemize}

\section{Methods}
\subsection{Problem Formulation}

Survival analysis aims to predict the risk of death according to gigapixel WSIs, which require huge computational resources. 
Limited by this, modern methods usually adopt a two-stage strategy that uses pre-trained networks to extract features from tiles and utilizes MIL algorithms to realize survival prediction by estimating hazard function $h(t)$. Given WSI $I_g$, $h(t)$ is represented as:
%$f_{hazard}(T=t|T\geq t, I_g)$,
% \begin{equation}
%     f_{hazard}(T=t) = \lim_{\Delta t \to 0} \frac{P(t\leq T \leq t + \Delta t | T \geq t)}{\Delta t},
% \end{equation}
\begin{equation}
    h(t)=P(T=t|T\geq t, I_g) = \lim_{\Delta t \to 0} \frac{P(t\leq T \leq t + \Delta t | T \geq t, I_g)}{\Delta t},
\end{equation}
which measures the instantaneous rate of occurrence of the event of interest at time $t$. $T$ is a variable of survival time. The survival function 
%$f_{surv}$, 
$S$,
% \begin{equation}
%     f_{surv}(T\leq t, I_g) = \prod^t_{u=1}(1-f_{hazard}(T=u)).
% \end{equation}
\begin{equation}
    S(t)=P(T\leq t|I_g) = \prod^t_{u=1}(1-h(u)),
\end{equation}
measures the probability of the patient surviving longer than $t$.
% In this paper, we build a novel end-to-end framework (VPTSurv) for survival analysis and compare it with the traditional end-to-end framework in Section \ref{sec:framework}.
% The details of VPTSurv are introduced in Section \ref{sec:vpt}, \ref{sec:mvpt}, and \ref{sec:decoder}, respectively.

\subsection{Two-stage Framework \textit{vs} End-to-end Framework}
\label{sec:framework}
\paragraph{\textbf{Two-stage:}}
To process gigapixel WSIs, existing methods typically slice a WSI $I_g$ into tens of thousands of patches $X=\{x_j|j\in [1, J]\}$ and then employ pre-trained networks such as ResNet~\cite{lu2021data} or ViT~\cite{huang2023visual} to extract feature vectors $V=\{v_j|j\in [1, J]\}$.
Then, $V$ is regarded as a bag to conduct multiple instance learning, or the patch vectors are formulated as the tokens of Transformer to model the sequence relations, to further estimate the hazard function.
Since the feature extractor is not optimized for survival analysis, the extracted features may lack important visual information essential for the task and thus may be unsuitable for survival analysis.
The key issues that limit the end-to-end training of survival models are: 1) Optimizing the feature extractor concurrently requires substantial GPU memory allocation, and 2) One WSI sample includes tens of thousands of patch images and demands GPU memory resources that exceed the capacity of current computing devices. To tackle these challenges, we propose a new framework to achieve end-to-end training for survival analysis in WSIs.

\begin{figure}
\includegraphics[width=\textwidth]{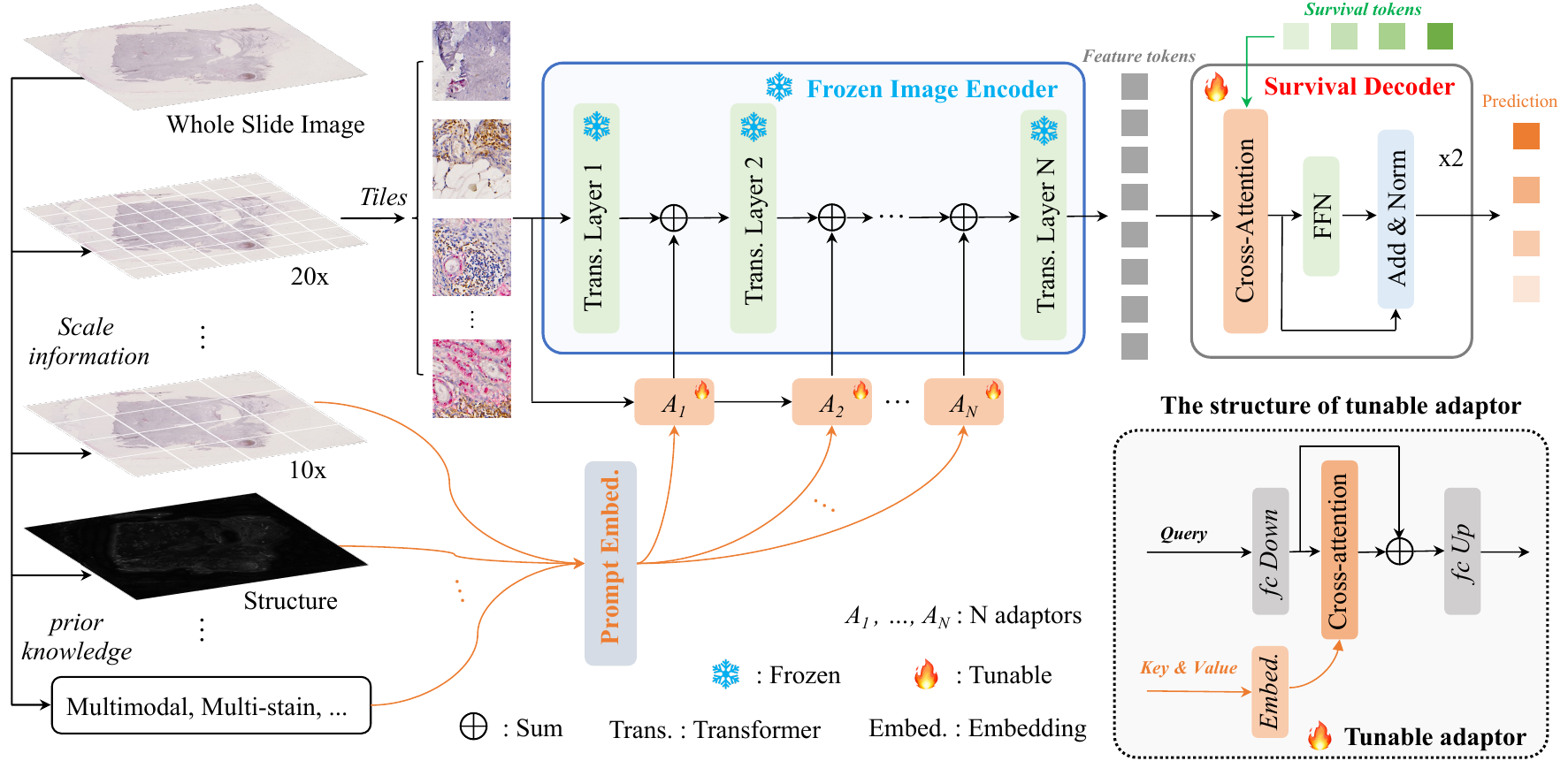}
\vspace{-0.8cm}
\caption{The framework of proposed VPTSurv in whole slide image.} \label{fig1}
\end{figure}

\paragraph{\textbf{End-to-end:}}
To achieve the end-to-end optimization for survival analysis in WSIs, we propose a multi-source Visual Prompt Tuning framework (VPTSurv). The overall framework of VPTSurv is shown in Figure \ref{fig1}.
VPTSurv consists of three parts: frozen image encoder, tunable adaptors, and survival decoder.
First, image patches $X=\{x_j|j\in [1, J]\}$ of size $H\times W$ are extracted from the input WSI $I_g$ in the scale level of $20\times$.
Then, $X$ is sent into the ViT-based image encoder $E$ to extract feature tokens. Image encoder includes $N$ Transformer layers, noted as $L = \{L_i|i\in [1,N]\}$.
$X$ is also sent into the $N$ adaptors to learn visual prompt tokens. Each adaptor corresponds to an encoder layer.
The encoder is frozen while adaptors are tunable during training. Thus, the extracted features can be tuned by adaptors.
Finally, the tokens from one patch are pooled into one token and further sent into the survival decoder $D$.
The cross-attention is computed between survival tokens and feature tokens to predict hazard function. The pipeline of VPTSurv can be formulated as:
% \begin{equation}
    $\tau_{hazard} = D(E(X, A(X)), \tau_{s})$,
% \end{equation}
where $A=\{A_i|i\in [1,N]\}$ represents the tunable adaptors. $\tau_{s}$ is the survival token queries and $\tau_{hazard}$ is the predicted survival tokens.

Compared with the traditional two-stage framework, VPTSurv solves the aforementioned two challenges.
First, VPTSurv introduces the visual prompt tuning into survival analysis, which enables the efficient tuning of image features by a few parameters in adaptors. The heavy feature extractor does not participate in the training, which speeds up the training and reduces the GPU memory footprint.
Second, VPTSurv architecture allows shorter sequence length during training while longer sequence length during inference due to its excellent ability of sequence length extrapolation, which comes from the combination of the proposed structure and Transformer.
During training, 200 patches are randomly selected and sent into the encoder to extract feature tokens. Then, the feature tokens from the same patches are pooled into one token. This operation decreases the huge number of tokens and avoids the softmax bottleneck in the cross-attention of the survival decoder. Moreover, during inference, the number of patches is variable, and more than 1000 patches can be utilized in VPTSurv without damaging performance (as no gradient computation is required).

\subsection{Visual Prompt Tuning for Survival Analysis}
\label{sec:vpt}
Visual Prompt Tuning (VPT) is a parameter-efficient finetuning scheme that embeds new visual prompts and inserts them into learnable adaptors, to further adaptively adjust the feature embedding from the pre-trained feature extractor for downstream survival analysis. Meanwhile, only the parameters of adaptors are learnable while the parameters of the backbone network are frozen. Therefore, VPT can adjust the extracted feature tokens by adaptive learning through the newly added visual prompts and adaptors, while preserving the capabilities of the original pre-trained network. The structure of VPT is shown in Figure \ref{fig1}. 
For a pre-trained feature extractor $E$, consisting of $N$ Transformer layers $L$,
\begin{equation}
    L_i(Q, K, V) = LM(FFN(softmax(\frac{Q\cdot K^{\top}}{\sqrt{d_k}})\cdot V + Q)) + Q,
\end{equation}
where $Q$, $K$, and $V$ are the query, key, and value tokens extracted from image patches $X$. $d_k$, $FFN$, and $LM$ represent the dimension of tokens, feed-forward network, and LayerNorm, respectively. One adaptor $A_i$ is configured for each layer within the encoder $E$. The visual prompts $Prompt_i$ can obtained as:
% \begin{equation}
\begin{align}
\label{eq:adaptor}
    Prompt_i &= A_i(Q^{'}, K^{'}, V^{'})= f_{up}(Att.(Q^{'}, K^{'}, V^{'}) + Q^{'}),\\
    Q^{'} &= f_{down}(Q),
    K^{'} = f_{emb}(K),
    V^{'} = f_{emb}(V),
\end{align}
% \end{equation}
where $f_{up}$, $f_{down}$, and $f_{emb}$ represent fully connected layers for upsampling, downsampling, and prompt embedding. $Att.(\cdot)$ represents a standard attention layer. $Q$, $K$, and $V$ represent the query, key, and value tokens. $Q$ originates from the visual prompt of the current patch image itself ($X$), while $K$ and $V$ can either come from the patch's visual prompts or other types of visual prompts.

Combined with the feature extractor, adaptors, and visual prompts, the new process of extractor feature tokens for survival analysis can be formulated as:
\begin{equation}
\label{eq:vpt}
    \tau_{i+1} = L_i(\tau_i) + A_i(\tau_i,\tau_i,\tau_i),
\end{equation}
where $\tau_i$ are the feature tokens in $i^{th}$ layer of the encoder and $\tau_{0}$ are tokens directly extracted from the tiles of by patch embedding of size $16 \times 16$.
VPTSurv is a general framework that can add more useful information to boost the accuracy of survival prediction.
The key and value tokens can come from tile images $X$ or other source information.
Based on this, we further propose multi-source visual prompt tuning for survival analysis in Section \ref{sec:mvpt}.

\subsection{Multi-source VPT for Survival Analysis}
\label{sec:mvpt}
The illustration of the multi-source VPT is shown in Figure \ref{fig1}. 
Assuming that there is some multi-source information like scale information $X_{scale}$, structural information $X_{structure}$, and prior knowledge $X_{prior}$ that could be useful for survival analysis, which can be denoted as $X_c = \{X_{scale}, X_{structure}, ..., X_{prior}\}$.
This information is used as the key and value token by the prompt embedding layer, to compute the cross-attention with original prompt tokens from patch images $X$. Then, the new prompts are embedded with multi-source information. According to Equation \ref{eq:adaptor} and \ref{eq:vpt}, the multi-source VPT can be formulated as:
\begin{equation}
    \tau_{i+1} = L_i (\tau_i) + A_i(\tau_i, f_{emb}(X_c),  f_{emb}(X_c)).
\end{equation}

In this paper, we introduce two useful information sources to boost the performance of survival analysis, which verifies the effectiveness of the multi-source VPTSurv framework.
First, due to the limited receive field on small tiles ($256\times 256$), the encoder only encodes the local cell information (cancer cells, immune cells) while lacking tissue information. Thus, we concatenate four neighboring tiles of $256\times 256$ to generate bigger patches of $512\times512$. Then, these bigger patches are resized into $256\times 256$ and further used as the scale visual prompt $X_{scale}$ for VPTSurv.
Second, the global structure information of WSIs is also important for the decision of survival analysis.
We also extract high-frequency components of WSIs as the structural visual prompt $X_{structure}$, formulated as:
\begin{equation}
    X_{structure} = \text{ifft}(\text{fft}(I_g) * mask),
\end{equation}
where fft and ifft represent the Fast Fourier Transform and its inverse. $mask$ is a binary mask to remove the low-frequency components.
Based on this framework, additional information such as multimodal information, multi-stain WSIs data, and prior knowledge about patients can all serve as prompts to enhance the performance of survival prediction.

\subsection{Survival Decoder}
\label{sec:decoder}
Most of the traditional two-stage methods focus on the second stage and design MIL algorithms to achieve survival prediction.
In the encoder-decoder framework, they can be regarded as the designs of the survival decoder.
They need to design complicated decoders to learn the useful features from the features extracted by the pre-trained network.
In our end-to-end VPTSurv framework, the feature tokens are well-optimized for survival prediction during training.
Therefore, the survival decoder is a simple network with two Transformer layers.
As shown in Figure \ref{fig1}, given the survival token queries $\tau_s$ and feature tokens $\tau_e$ from the encoder, the cross-attention is computed with $\tau_s$ and $\tau_e$ to output the survival prediction $\tau_{hazard}$.
% \subsection{Loss Function}
% \label{loss}
Following~\cite{zadeh2020bias}, the negative log-likelihood loss is used. $\tau_{hazard}$ is used as predicted hazard function $h(t)$ to compute survival function $S(t)$. Finally, the loss $\mathcal{L}$ is:
\begin{align}
\mathcal{L} = & -\sum_{(X,t,c)\in \mathcal{D}_{train}} c \cdot \text{log}(S(t|X)) \nonumber\\
&- \sum_{(X,t,c)\in \mathcal{D}_{train}}\{(1-c)\cdot \text{log}(S(t-1)|X)+(1-c)\cdot \text{log}(h(t|X))\},
\end{align}
where $(X,t,c)$ represents the WSI, survival time, and the state of right-censored for patients that sampled from dataset $\mathcal{D}$.

\section{Experiments}
\paragraph{\textbf{Datasets and Metric:}}

To evaluate the effectiveness of the proposed VPTSurv on survival analysis, we collect two pancreatic cancer datasets: IHC-CD4 and IHC-CD8. The IHC-CD4 dataset includes 564 patients with 655 WSI images, which are randomly divided into the train (449), validation (71), and test (135) sets. The IHC-CD8 dataset includes 564 patients with 754 WSI images, which are randomly divided into train (509), validation (83), and test (162) sets.
For evaluation, the concordance index (CI) is used to verify the performance of the proposed methods and compare them with other methods. VPTSurv is trained on the training set and the best models are selected on the validation set according to CI. The results are reported on the test set.

\paragraph{\textbf{Implementation Details:}}
The frozen image encoder is ViT-base pre-trained by \cite{Kirillov_2023_ICCV}, which includes 12 Transformer layers. The adaptor number is 12. The dimension of feature tokens is 768, which is decreased to 24 by $f_{down}$ and increased to 768 by $f_{up}$. The decoder layer number and dimension are 2 and 768, respectively. The patch numbers for training and inference are 200 and 1000. The optimizer is AdamW, and the learning rate is 5e-5 with a CosineAnnealingLR scheduler. The models are trained with 15 epochs on 8 Nvidia A800 GPUs.

\begin{table}[]

\centering
\caption{The main results of VPTSurv on IHC-CD4 and IHC-CD8 datasets.}
\vspace{-0.3cm}
\renewcommand\arraystretch{1.1}
\renewcommand\tabcolsep{4pt}
\begin{tabular}{|c|c|c|c|}
\hline

\hline
Model & Category & IHC-CD4 (CI \%) & IHC-CD8 (CI \%) \\
\hline

\hline
TransMIL~\cite{shao2021transmil} & \multirow{4}{*}{Two-stage} & 51.75 & 56.11\\
PatchGCN~\cite{chen2021whole} &  & 54.01 & 51.93\\
DeepAttnMISL~\cite{yao2020whole} & & 56.55 & 55.57 \\
HVTSurv~\cite{shao2023hvtsurv} & & 56.23 & 52.17\\
\hline
Only VPTSurv Decoder (Ours) & Two-stage & 55.12 & 54.32 \\
\hline
VPTSurv (Ours) &  End-to-end & 59.93 & 61.10\\
\hline

\hline
\end{tabular}
\label{tab:1}
\end{table}

\vspace{-1cm}

\begin{figure}
\includegraphics[width=\textwidth]{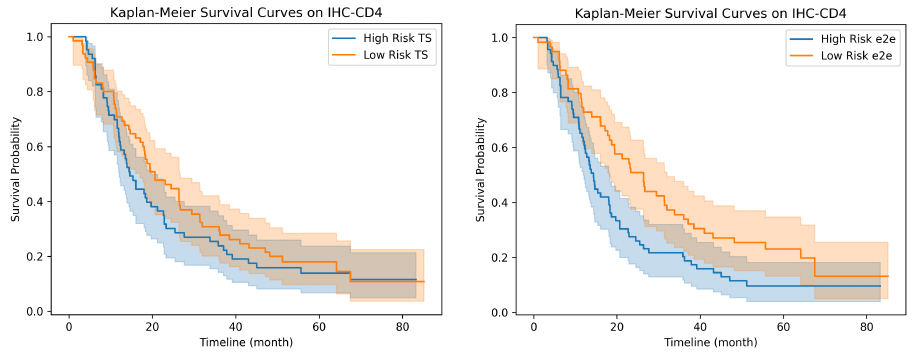}
\vspace{-0.9cm}
\caption{Kaplan-Meier analysis of two-stage (TS) and end-to-end (e2e) framework.} \label{fig2}
\end{figure}

\paragraph{\textbf{Main Results:}}
VPTSurv is evaluated on two datasets with different immunohistochemical pathology images in Table \ref{tab:1}. Compared with typical two-stage methods such as TransMIL~\cite{shao2021transmil}, PatchGCN~\cite{chen2021whole}, DeepAttnMISL~\cite{yao2020whole}, and HVTsurv~\cite{shao2023hvtsurv}, we obtain 55.12\% and 54.32\% in C-index with only using VPTSurv decoder, which is comparable or slightly below these two-stage approaches.
However, VPTSurv achieves 59.93\% and 61.10\% in C-index under the end-to-end scheme, with significant relative improvements of 8.7\% and 12.5\% on IHC-CD4 and IHC-CD8, respectively. We also visualize the Kaplan-Meier survival curves in Figure \ref{fig2}. These results show that the survival prediction of end-to-end VPTSurv shows a clearer grouping of high and low risk than the two-stage method.

\paragraph{\textbf{Ablation Study:}}
VPTSurv includes the frozen ViT encoder, adaptors for visual prompt tuning, and survival decoders.
Since the existing method usually adopts the two-stage scheme, following this, we build a two-stage baseline with only the VPTSurv decoder. 
As shown in Table \ref{tab_ab}, using all patches whose features are extracted from a pre-trained network, the two-stage method achieves a CI of 54.58\%.
Then, add the frozen image encoder and visual prompt tuning VPTSurv and make it end-to-end trainable. VPTSurve achieves a CI of 58.56\%, with a relative improvement of 7.5\% compared with the two-stage approaches.
These results demonstrate the significant advantages of the end-to-end scheme compared to the two-stage strategy.
To evaluate the feasibility of the multi-source VPTSurv, two additional visual prompts are added to boost the performance. As shown in Table \ref{tab_ab}, by adding structural and scale prompts, VPTSurv achieves a CI of 58.85 and 59.93, outperforming the VPTSurv baseline without using additional prompts.
These results demonstrate that the multi-source VPT architecture is effective and scalable. Using structural and multi-scale information as prompts is beneficial for prognosis analysis.

\begin{table}[]

\centering
\caption{Ablation study of VPTSurv on IHC-CD4 dataset.}
\vspace{-0.3cm}
\renewcommand\arraystretch{1.1}
\renewcommand\tabcolsep{4pt}
\begin{tabular}{|c|c|c|c|c|}
\hline

\hline
Model & Category & Feature & CI (\%) \\
\hline

\hline
Only VPTSurv Decoder &  Two-stage & - & 55.12\\
\hline
VPTSurv &  End-to-end & VPT & 58.56 \\
VPTSurv &  End-to-end & VPT + structure & 58.95 \\
VPTSurv &  End-to-end & VPT + structure + scale & 59.93\\
\hline

\hline
\end{tabular}\label{tab_ab}
\end{table}

\vspace{-1cm}
\section{Discussion and Conclusion}
VPTSurv is the pioneering end-to-end VPT framework for survival analysis. It presents a viable architecture capable of incorporating additional modalities, textual sources, or tabular data as prompts to enhance the accuracy of survival predictions. Exploring these extensions will constitute our future work.

In this paper, we introduce the first end-to-end multi-source VPT network for survival analysis in WSIs. To overcome the challenges associated with end-to-end training, we employ VPT to efficiently learn relevant features for survival analysis. This approach significantly reduces GPU memory requirements, facilitating the training of survival models on hundreds of tiles and allowing inference on thousands of tiles. 
VPTSurv demonstrates that end-to-end training can significantly enhance the accuracy of prognostic analysis.
Furthermore, we show that incorporating structural and scale information as additional prompts in VPTSurv can substantially improve the accuracy of survival predictions.

\clearpage
\bibliographystyle{splncs04}
\bibliography{ref}




% LNCS Homepage, \url{http://www.springer.com/lncs}. Last accessed 4
% \end{thebibliography}
\end{document}